\documentclass[11pt]{article}

\usepackage{natbib}

\usepackage{graphicx}
\usepackage{here}
\usepackage[labelfont=bf,hang]{caption}
\usepackage[labelfont=bf]{subfig}
\usepackage{color}
\usepackage{bm}
\usepackage{amsmath, amsthm, amsfonts}

\usepackage{cases}
\usepackage{newtxtext}
\usepackage{fancyhdr}
\usepackage{longtable}
\usepackage{here}

\usepackage{enumitem}
	\setlist[itemize]{noitemsep}
	\setlist[description]{noitemsep}
	\setlist[enumerate]{noitemsep, font=\bf}

\usepackage{endnotes}

\usepackage[top=30truemm,bottom=30truemm,left=30truemm,right=30truemm]{geometry}

\usepackage[
	bookmarks=false,
	colorlinks=true,
	citecolor=blue,
	urlcolor=blue,
	linkcolor=blue,
 	pdfstartview=FitH,
]{hyperref}

\usepackage[capitalise]{cleveref}
\crefname{theo}{Theorem}{Theorems}
\crefname{prop}{Proposition}{Proposition}
\crefname{eq}{Eq.}{Eqs.}
\crefname{fig}{Fig.}{Figs.}
\crefname{tab}{Tab.}{Tabs.}
\crefname{sec}{Section}{Sections}

\usepackage{listings,color}
\definecolor{OliveGreen}{rgb}{0.0,0.6,0.0}
\definecolor{Orenge}{rgb}{0.89,0.55,0}
\definecolor{SkyBlue}{rgb}{0.28, 0.28, 0.95}
\def\({\left(}
\def\){\right)}
\def\[{\left[}
\def\]{\right]}
\def\<{\left\langle}
\def\>{\right\rangle}
\def\<<{\left\{}
\def\>>{\right\}}

\def\Dn{\Delta n}

\def\->{$\rightarrow$}

\allowdisplaybreaks

\reversemarginpar

\begin{document}


\title{UXsim: An open source macroscopic and mesoscopic traffic simulator in Python---a technical overview}
\author{
	Toru Seo\footnote{Tokyo Institute of Technology; seo.t.aa@m.titech.ac.jp; The corresponding author}
}

\date{\today}
\maketitle

\begin{abstract}
	This note describes a technical overview of UXsim, an open source macro/mesoscopic traffic simulator in pure Python programming language.
	UXsim is based on Kinematic Wave model (more specifically, mesoscopic version of Newell's simplified car-following model) and dynamic user optimum-like route choice principle, which are well established methodology in the transportation research field.
	It can compute dynamical network traffic flow and have basic visualization and analysis capability.
	Furthermore, users can implement their own models and control methods into the simulator by using Python, thanks to the flexibility of the language.
	The simulator and its codes are freely available at \url{https://github.com/toruseo/UXsim} under the MIT license.
\end{abstract}




\section{Introduction}

Vehicular traffic flow plays essential roles in today's civilization.
However, it faces several critical issues such as congestion, accidents, and environmental burden.
Macroscopic traffic simulation is important to understand and manage dynamic urban-scale vehicular traffic flow.
Mesoscopic traffic simulation is a type of macroscopic traffic simulation, in which some microscopic nature of traffic is incorporated in order to enhance simulation capability while keeping computational efficiency high.

{\it UXsim} is a new open source macroscopic and mesoscopic traffic simulator developed by the author.
It is written in pure Python and its common libraries.
Thus, it can be used and modified flexibly by users.
Although Python's computational efficiency is not so high, UXsim can compute large-scale traffic phenomena fairly effectively thanks to the use of the macro/mesoscopic model.

This note describes a technical overview of UXsim, namely, its simulation logic with simple examples.
The details are throughly explained in \citet{seo2023book_en}, a Japanese book on general macroscopic traffic flow theory and simulation.
The main functions of UXsim are as follows:
\begin{itemize}
	\item Dynamic network traffic simulation with a given network and time-dependent OD demand.
	\item Implementation of traffic management schemes (e.g., traffic signals, inflow control, route guidance, congestion pricing).
	\item Basic analysis of simulation results (e.g., trip completion rate, total travel time, delay), and their export to {\tt pandas.DataFrame} and CSV files.
	\item Visualization of simulation results (e.g., time-space diagram, MFD, network traffic animation).
\end{itemize}

Just for information, the origin of the name ``UXsim'' is as follows.
``U'' stands for uroboros, a mythical snake that embodies some essence of network traffic congestion.
``X'' means the position coordinates, which is the most important state variable in the employed mesoscopic traffic flow model.
``sim'' signifies, of course, simulation.
UXsim and its codes are freely available at \url{https://github.com/toruseo/UXsim} under the MIT license.

\section{Simulation logic}

\subsection{Models}

\subsubsection{Overview}\label{sec_over}

The simulator is based on Kinematic Wave (KW) model \citep{Lighthill1955LWR,Richards1956LWR} and dynamic user optimum-like route choice principle, which are well established and common methodology in the transportation research field.
More specifically, it is constructed by combining the following models.

Link/Vehicle model determines traffic dynamics on links.
The mesoscopic version of Newell's simplified car-following model \citep{newell2002carfollowing}, also known as X-model \citep{Laval2013trafficflow} is employed for this purpose.
This is equivalent to KW model with a triangular fundamental diagram.
In this model, vehicles travels as fast as possible, while maintaining safe spacing and headway which are speed-dependent.
Thus, it can reproduce traffic congestion and queuing phenomena fairly accurately.

Node model determines inter-link transfer of traffic.
The mesoscopic version of the incremental node model \citep{flotterod2011node, flotterod2016node}  is employed for this purpose.
This model can reproduce merging and diverging traffic in a manner consistent with KW model.

Route choice model determines network-level decision of travelers.
Dynamic user optimum \citep{Kuwahara2001duo}, also known as the reactive assignment, with stochasticity and delay is employed for this purpose.
In this model, travelers tend to choose the shortest path to the destination that minimizes instantaneous travel time.
However, instantaneous travel time is highly volatile, so a kind of inertia is added to traveler's decision making process by incorporating stochasticity and delay.

\subsubsection{Details}

For reference, important model formula are presented in this section.
Users do not necessarily need to understand these contents to use UXsim, but understanding on them would be useful for advanced usage and customization of UXsim.
For further details, please refer to \citet{seo2023book_en} or the original articles mentioned in \cref{sec_over}.

Regarding the Link/Vehicle model, the driving behavior of a platoon consists of $\Delta n$ vehicles in a link is expressed as
\begin{align}
    X(t+\Delta t,n) = \min\left\{\begin{array}{l}
        X(t, n) + u\Delta t,    \\
        X(t +\Delta t-\tau\Dn, n-\Dn) - \delta\Dn
    \end{array}\right\},    
\end{align}
where $X(t,n)$ denotes position of platoon $n$ on time $t$, $\Delta t$ denotes simulation time step width, $u$ denotes free-flow speed of the link, $\delta$ denotes jam spacing of the link.

The Node model is computed by the following algorithm:
\begin{enumerate}
    \item Let $i$ be time step number
    \item Select incoming link $l$ with probability $\alpha_l/\sum_l \alpha_l$, where $\alpha_l$ is the merging priority parameter of link $l$.
    	If all incoming link have been selected, go to step 6.
    \item Select the vehicle that exists at the end of link $l$.
    	Hereafter, the vehicle is denoted as $n$.
    	If there is no such vehicle, go back to step 2.
    \item Let $o_n^i$ be the outgoing link that vehicle $n$ want to go.
    	Check whether link $o_n^i$ has a sufficient (larger than $\delta\Dn$) vacant space at its starting position, and do the following:
    \begin{itemize}
        \item If yes, transfer vehicle $n$ from link $l$ to link $o_n^i$.
        \item If no, vehicle $n$ cannot move, so do nothing.
    \end{itemize}
    \item Go back to step 2.
    \item Increment time step ($i:=i+1$) and go back to step 2.
\end{enumerate}

The Route choice model is computed by the following steps.
Let $b_o^{z,i}$ be a dummy variable that is 1 if link $o$ is a part of the shortest path from every nodes to destination $z$ on time step $i$, and 0 otherwise.
Shortest path search is performed every $\Delta i_B$ time steps.
When shortest path search is performed, we update $B_o^{z,i}$, an attractiveness of link $o$ for vehicles with destination $z$ on time step $i$, as 
\begin{align}
    B_o^{z,i} = (1-\lambda) B_o^{z, i-\Delta i_B} + \lambda b_o^{z,i},
\end{align}
where $\lambda$ is a given weight.
We assume that the initial $B_o^{z,0}$ is equal to $b_o^{z,0}$.
Finally, outgoing link $o^i_n$ of vehicle $n$ is determined as $o$ with probability $B_o^{z,i}/\sum_o B_o^{z,i}$, where $z$ is the destination of $n$.

\subsection{Implementation}

\begin{figure}[htp]
	\centering
	\includegraphics[clip, width=0.99\hsize]{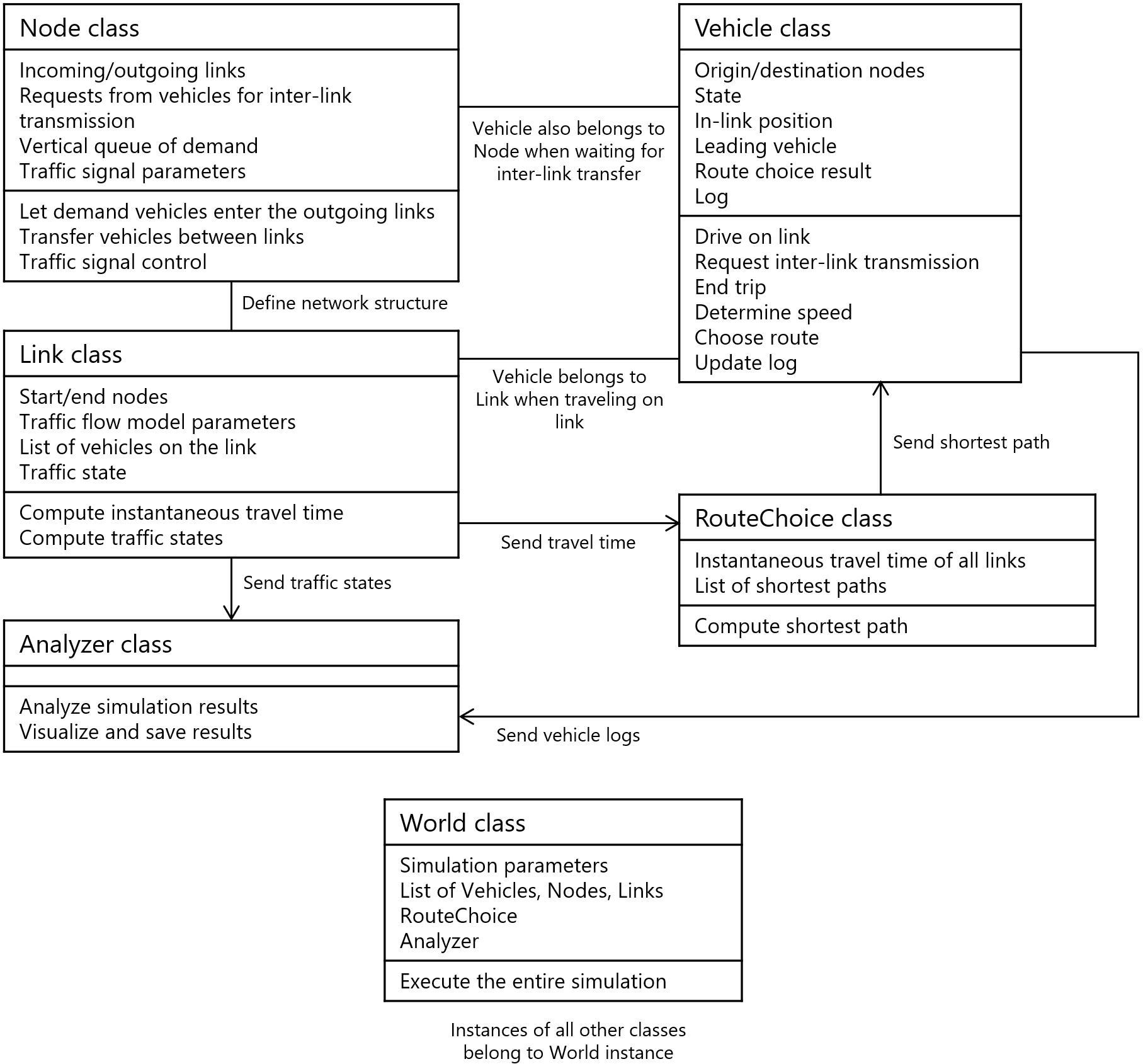}
	\caption{Class diagram of UXsim (translated and modified from \citet{seo2023book_en})}
	\label{class_diagram}
\end{figure}

The implementation of the models are summarized in the following diagrams. 
\cref{class_diagram} shows a static structure of UXsim as a class diagram (based roughly on Unified Modeling Language).
Each three-row rectangle represents a class.
The top row denotes the class name, the second row denotes key variables, and the third row denotes key functions.

\begin{figure}[htp]
	\centering
	\subfloat[Entire simulation]{\includegraphics[clip, width=0.95\hsize]{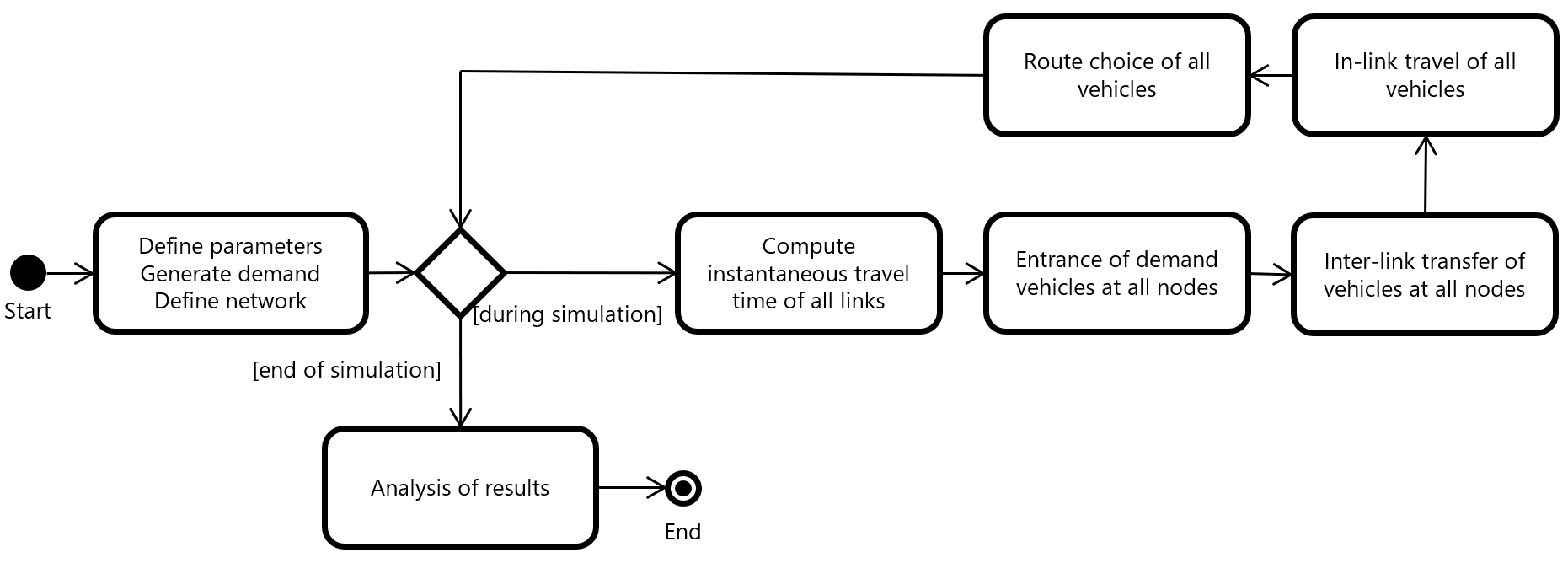}\label{activity_diagram}}\\
	\subfloat[Specific vehicle]{\includegraphics[clip, width=0.8\hsize]{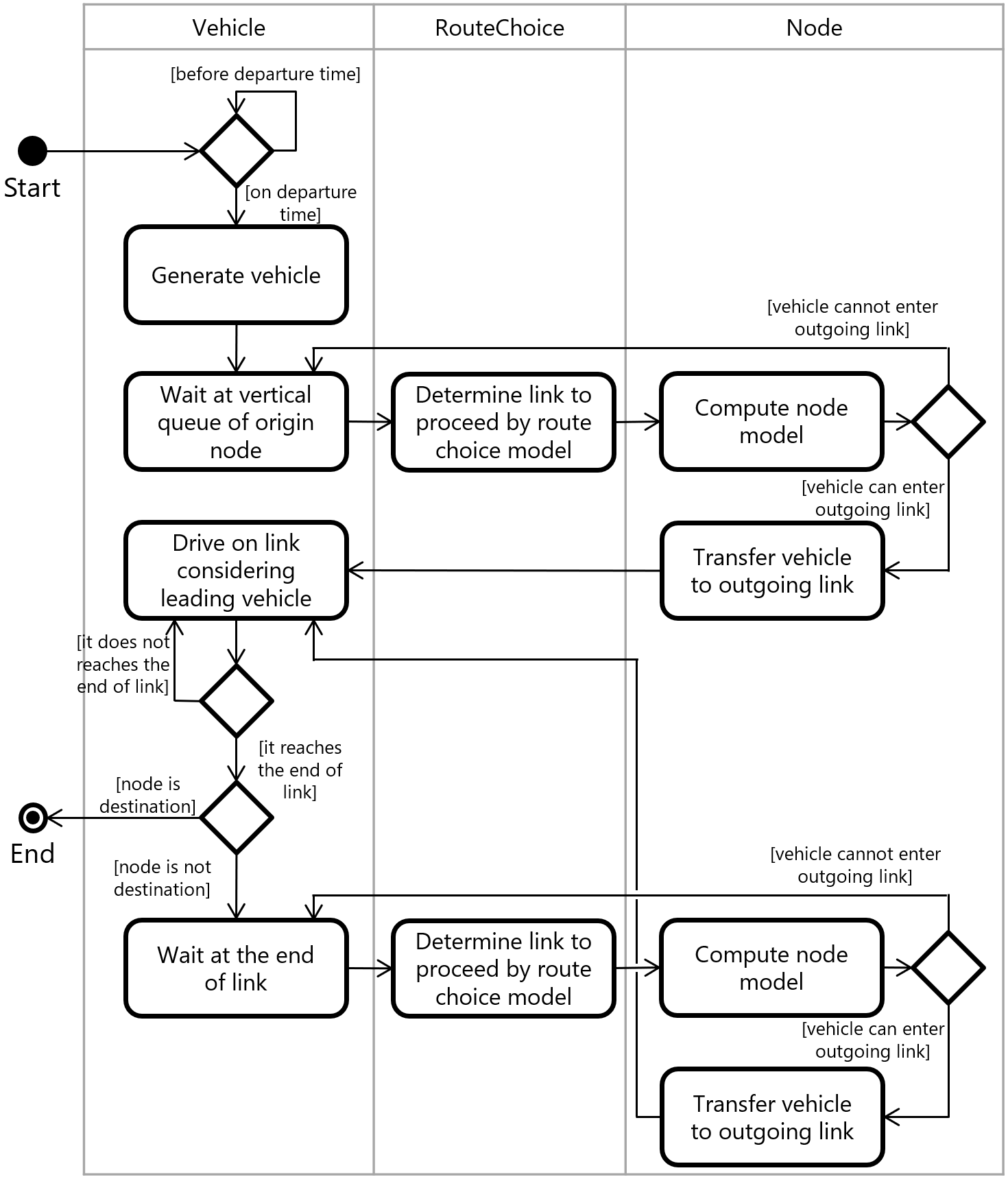}\label{activity_diagram_veh}}
	\caption{Activity diagrams of UXsim (translated and modified from \citet{seo2023book_en})}
	\label{activity_diagram0}
\end{figure}

Similarly, \cref{activity_diagram} shows a dynamic computational flow of entire simulation of UXsim as an activity diagram.
\cref{activity_diagram_veh} shows that of an instance of Vehicle class in UXsim.
For the details, please see the codes.

\subsection{Key inputs}

Some of the key inputs for simulation are as follows:
\begin{itemize}
	\item Reaction time of vehicles $\tau$. 
		This value determines the simulation time step width.
		This is a global parameter.
	\item Platoon size for mesoscopic simulation $\Delta n$.
		This value determines the simulation time step width.
		This is a global parameter.
	\item Route choice model parameters: shortest path update interval $\Delta i_B$ and weight value $\lambda$.
	\item Lists of nodes and links.
		They define the network structure.
	\item Parameters of each link.
		For example, length, free flow speed $u$, jam density $\kappa$, and merging priority parameter $\alpha$.
	\item Parameters of each node.
		For example, position and traffic signal setting.
	\item Demand.
		For example, origin, destination, and departure time of each vehicle are specified.
\end{itemize}
Roughly estimating, the computational cost of the simulation is proportional to the total number of vehicles, and inverse proportional to $\tau$ and $\Delta n^2$.

\subsection{Analysis}

UXsim has several built-in analysis and visualization functions.
Examples are as follows:
\begin{itemize}
	\item Overall traffic analysis (e.g., total travel time)
	\item OD-level traffic analysis (e.g., travel time, delay)
	\item Link-level traffic analysis (e.g., traffic volume, delay, dynamic traffic states)
	\item Output simulation and analysis results to file or pandas dataframe
	\item Time-space diagram of trajectories of each link
	\item Time-space diagram of traffic states (i.e., flow, density, and speed) of each link
	\item Cumulative counts of each link
	\item Macroscopic fundamental diagram (MFD) \citep{Geroliminis2007mfd}
	\item Animation of network traffic state dynamics
	\item Animation of vehicle flow in network
\end{itemize}

\section{Examples}

\subsection{Gridlock and its prevention}

\begin{figure}[htp]
	\centering
	{\includegraphics[clip, width=0.65\hsize]{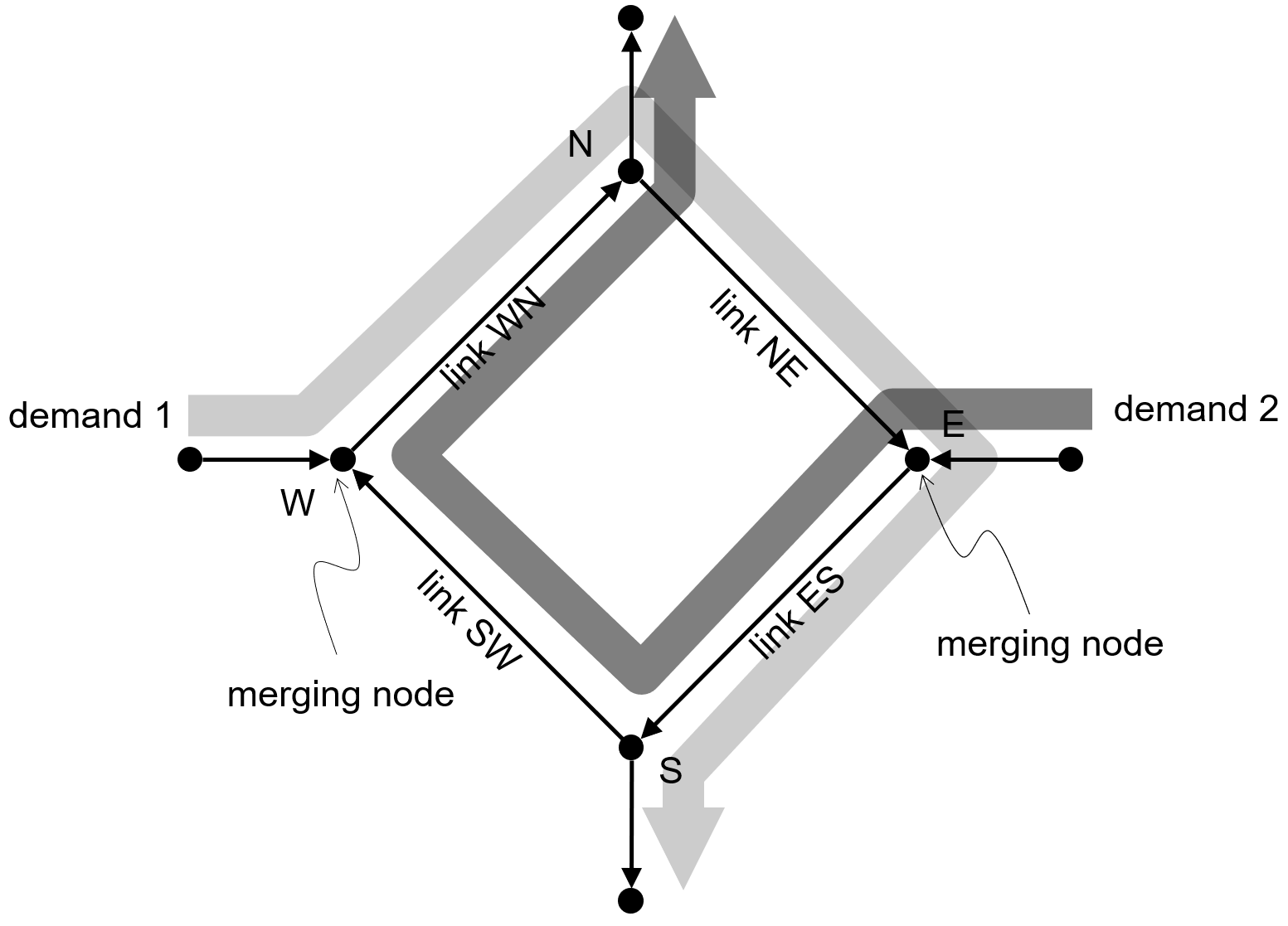}\label{uroboros_net}}
	\caption{Gridlock scenario (translated and modified from \citet{seo2023book_en})}
	\label{scenario_gridlock}
\end{figure}

Simulation of a simple gridlock congestion is shown as an example.
The scenario setting is illustrated in \cref{scenario_gridlock}.
There is one circular road, and the demands are arranged in such a way that they interfere with each other.
In this way, the beginning of some traffic jams and the end of others may engage, resulting in a gridlock condition.

\begin{figure}[htp]
	\centering
	\subfloat[{\tt uroboros\_nodes.csv}]{\includegraphics[clip, width=0.3\hsize]{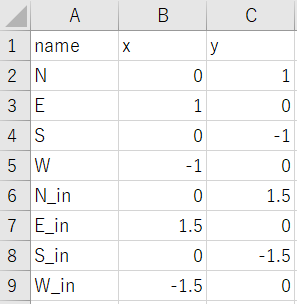}\label{uro_nodes}}~~~~
	\subfloat[{\tt uroboros\_links.csv}]{\includegraphics[clip, width=0.5\hsize]{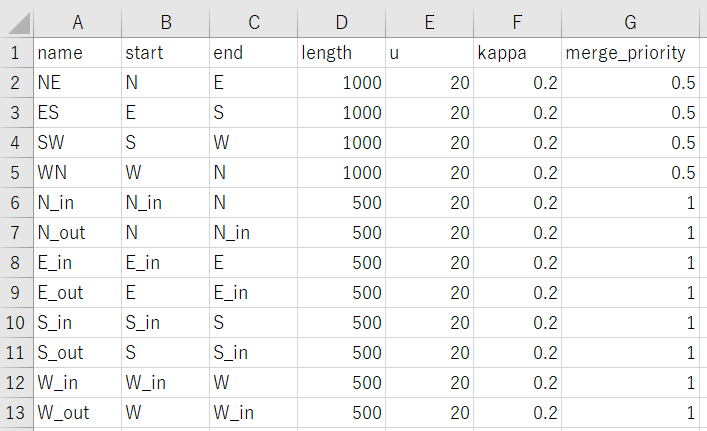}\label{uro_links}}\\
	\subfloat[{\tt uroboros\_demand.csv}]{\includegraphics[clip, width=0.4\hsize]{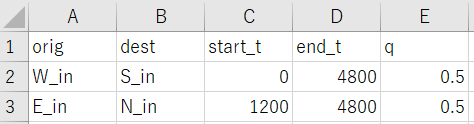}\label{uro_demand}}
	\caption{Contents of CSV files defining scenario}
	\label{scenario_csv}
\end{figure}

Such simulation can be executed by Code \ref{gridlock_code}.
The contents of each CSV file loaded at lines 16 and 17 are shown in \cref{scenario_csv}.
These code and CSV files are included in the repository as {\tt demos\_and\_examples/example\_05\_gridlock\_and\_prevention.py}.

\begin{lstlisting}[caption=Gridlock, label=gridlock_code, language=Python]
from uxsim import *
import pandas as pd

if __name__ == "__main__":
    # Define simulation
    W = World(
        name="",
        deltan=5,
        print_mode=1, save_mode=1, show_mode=0,
        random_seed=0
    )
    
    # Define scenario
    # import CSV files
    W.load_scenario_from_csv("dat/uroboros_nodes.csv", 
        "dat/uroboros_links.csv", "dat/uroboros_demand.csv")

    W.finzalize_scenario()
    
    # Execute simulation
    W.exec_simulation()
    
    # Result analysis and save
    W.analyzer.print_simple_stats()
    W.analyzer.output_data()
    W.analyzer.time_space_diagram_traj_links([["WN", "NE", "ES", "SW"]])
    W.analyzer.macroscopic_fundamental_diagram()
    W.analyzer.network_anim(detailed=1, network_font_size=0, figsize=(12,12))
    W.analyzer.network_fancy(animation_speed_inverse=15, sample_ratio=0.3, interval=5, trace_length=5)
    
    df_vehicles = W.analyzer.log_vehicles_to_pandas()
\end{lstlisting}

\begin{figure}[htp]
	\centering
	\subfloat[Vehicle trajectories in circular road]{\includegraphics[clip, width=0.99\hsize]{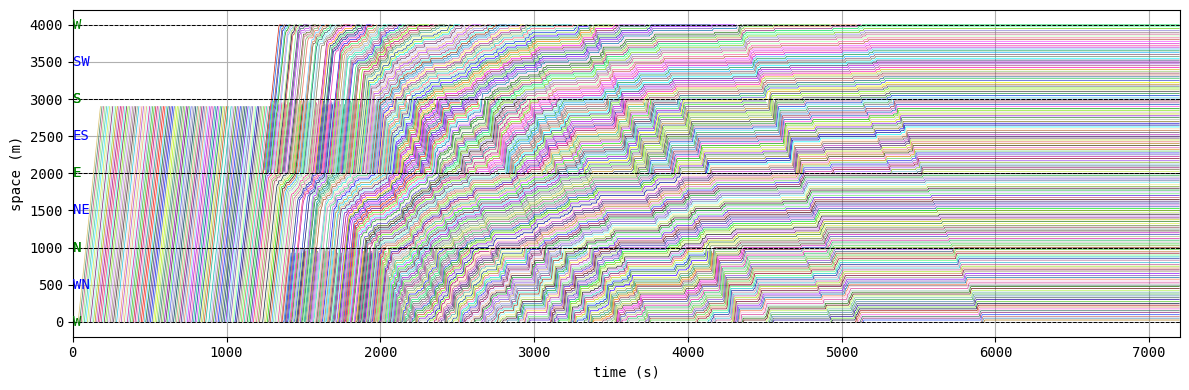}\label{tsd_gridlock}}\\
	\subfloat[MFD]{\includegraphics[clip, width=0.3\hsize]{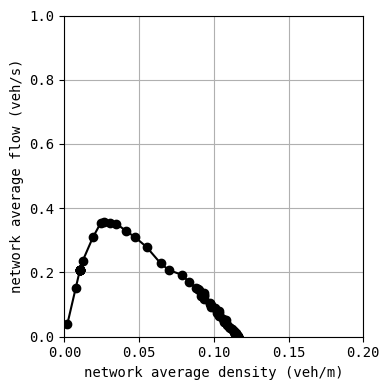}\label{mfd_gridlock}}
	\caption{Gridlock simulation}
	\label{res_gridlock}
\end{figure}

Some of the outputs of this simulation are shown in \cref{res_gridlock}.
According to vehicle trajectories shown in \cref{tsd_gridlock}, we can see that traffic was free-flowing until 1200 s when only one direction traffic demand was traveling.
However, after the second demand started traveling, traffic congestion immediately occurred at the merging nodes W and E.
Each queue generated by traffic congestion extended quickly and reaches the heads of the other queue.
Then, gridlock happened.
The MFD in \cref{mfd_gridlock} clearly shows a typical gridlock phenomenon.
Other outputs such as gif animations are also useful to confirm the dynamics of gridlock phenomena.

Now let us consider how to prevent a gridlock by implementing proper traffic management.
For example, because the merging nodes are the sources of the queues that triggered the gridlock, we can prevent the gridlock by increasing the merging priority of links in the circular road.
This can be done by inserting Code \ref{gridlock_code2} to line 17 of Code \ref{gridlock_code}: here, the merging priority parameters of links NE and SW are increased from the default value 0.5 to 2
In the real world, this kind of management can be executed by doing ramp metering, signal control, and perimeter control.
\begin{lstlisting}[caption=Gridlock prevention management, label=gridlock_code2, language=Python]
    W.get_link("NE").merge_priority = 2
    W.get_link("SW").merge_priority = 2
\end{lstlisting}

\begin{figure}[htp]
	\centering
	\subfloat[Vehicle trajectories in circular road]{\includegraphics[clip, width=0.99\hsize]{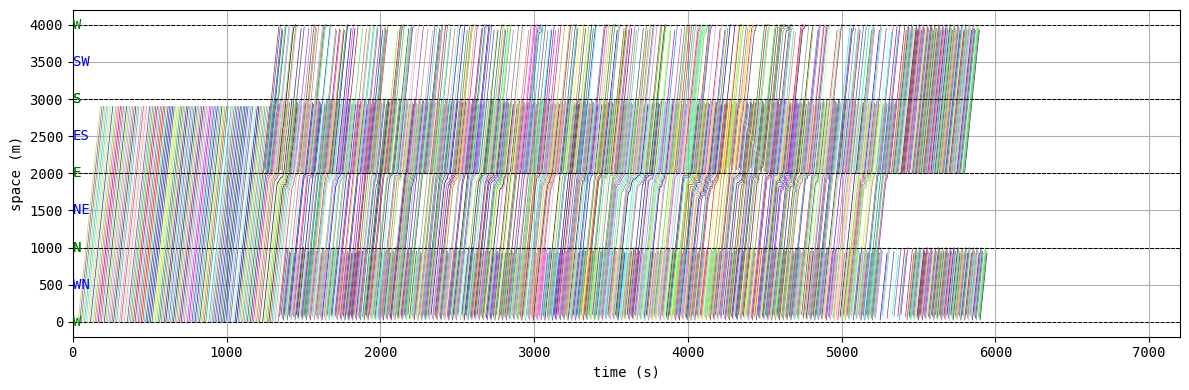}\label{tsd_manage}}\\
	\subfloat[MFD]{\includegraphics[clip, width=0.3\hsize]{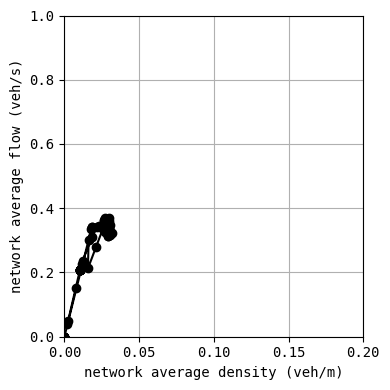}\label{mfd_manage}}
	\caption{Simulation that prevented gridlock}
	\label{res_manage}
\end{figure}

The results are shown in \cref{res_manage}.
It is obvious that gridlock is perfectly prevented.
The circular road was almost always free-flowing, and the MFD's state was also in free-flowing or critical regime.
Although some congestion occurred at the entry links (this can confirmed by analyzing other outputs), such congestion quickly diminishes one traffic in the circular road finishes traveling.
Thus, this kind of traffic management is very effective to prevent traffic congestion and gridlock with little or no loss to anyone.

\subsection{Sioux Falls network}

\begin{figure}[htp]
	\centering
	\subfloat[Network traffic state. The width of each link denotes the number of vehicles on the link, and the color of each link denotes its average speed (the darker the slower).]{\includegraphics[clip, width=0.75\hsize]{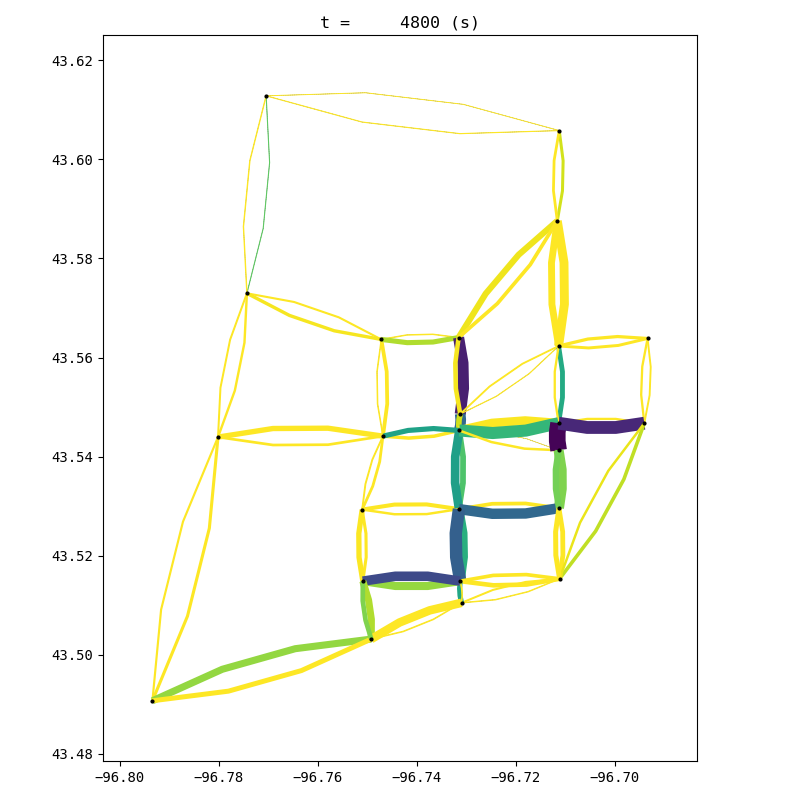}\label{sf_net_states}}\\
	\subfloat[Vehicle trajectories in a corridor at the center of network, northbound direction.]{\includegraphics[clip, width=0.99\hsize]{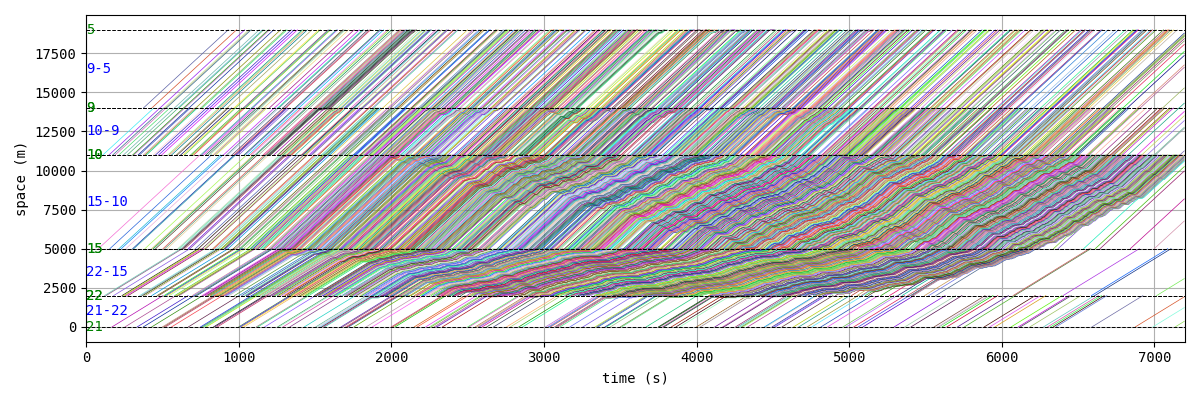}\label{sf_net_traj}}
	\caption{Simulation results in Sioux Falls network}
	\label{sf_net}
\end{figure}

Sioux Falls network is a popular network for testing and benchmarking in the transportation research field \citep{tnrct2021network}.
\cref{sf_net} shows simulation results by UXsim.
A qualitatively reasonable traffic pattern on the network was obtained by considering traffic dynamics and route choice.
The specification of the simulation scenario is as follows:
\begin{itemize}
	\item Simulation duration: 7200 s
	\item Total number of vehicles: 34690 veh
	\item Number of links: 76
	\item Total road length: 314000 m
	\item Time step width $\Delta t$: 5 s
	\item Platoon size $\Delta n$: 5 veh
	\item Route choice update interval $\Delta i_B \Delta t$: 600 s
\end{itemize}
The computation time was about 16 s using Windows 10 computer with 3.79 GHz CPU and 32 GB RAM.

The repository \url{https://github.com/toruseo/UXsim} also contains detailed other examples such as highway bottleneck, traffic signal control, scenario generation by script, along with detailed documents.


\end{document}